\documentclass[aps,twocolumn,superscriptaddress]{revtex4-1}
\usepackage[colorlinks=true,bookmarks=true,citecolor=magenta,urlcolor=magenta,linkcolor=magenta,breaklinks]{hyperref} 
\usepackage{breakurl}
\usepackage{sidecap}
\usepackage{amssymb}
\usepackage{hhline}
\usepackage{urwchancal}
\usepackage{xcolor}
\sidecaptionvpos{figure}{t}
\usepackage{amsmath}
\usepackage{graphicx}
\usepackage{esint}
\usepackage{epstopdf}
\usepackage{rotating}
\graphicspath{{pict/}{./}}
\usepackage{bm}%
\usepackage{microtype,bm,bbm,graphicx,booktabs,times}

\newcounter{Fig}

\begin{document}


\title{Symmetry Protected Invariant Scattering Properties for Arbitrary Polarizations}
\author{Qingdong Yang}
\email{Authors contributed equally to this work.}
\affiliation{School of Optical and Electronic Information, Huazhong University of Science and Technology, Wuhan, Hubei 430074, P. R. China}
\author{Weijin Chen}
\email{Authors contributed equally to this work.}
\affiliation{School of Optical and Electronic Information, Huazhong University of Science and Technology, Wuhan, Hubei 430074, P. R. China}
\author{Yuntian Chen}
\email{yuntian@hust.edu.cn}
\affiliation{School of Optical and Electronic Information, Huazhong University of Science and Technology, Wuhan, Hubei 430074, P. R. China}
\affiliation{Wuhan National Laboratory for Optoelectronics, Huazhong University of Science and Technology, Wuhan, Hubei 430074, P. R. China}
\author{Wei Liu}
\email{wei.liu.pku@gmail.com}
\affiliation{College for Advanced Interdisciplinary Studies, National University of Defense
Technology, Changsha, Hunan 410073, P. R. China}

\begin{abstract}
Polarization independent Mie scattering of building blocks is foundational for constructions of optical systems with robust functionalities. Conventional studies for such polarization independence are generally restricted to special states of either linear or circular polarizations, widely neglecting elliptically-polarized states that are generically present in realistic applications. Here we present a comprehensive recipe to achieve invariant scattering properties (including extinction, scattering and absorption) for arbitrary polarizations, requiring only rotation symmetry and absence of optical activities.  It is discovered that sole rotation symmetries can effectively decouple the two scattering channels that originate from the incident circularly polarized waves of opposite handedness, leading to invariance of all scattering properties for any polarizations on the same latitude circle of the Poincar\'{e} sphere. Further incorporations of extra inversion or mirror symmetries would eliminate the optical activities and thus ensure scattering property invariance for arbitrary polarizations. In sharp contrast to previous investigations that rely heavily on complicated algebraic formulas, our arguments are fully intuitive and geometric, bringing to surface the essential physical principles rather than obscuring them. The all-polarization invariance we reveal is induced by discrete spatial symmetries of the scattering configurations, underlying which there are functioning laws of reciprocity and conservation of parity and helicity.  This symmetry-protected intrinsic invariance is robust against any symmetry-preserving perturbations, which may render extra flexibilities for designing optical devices with stable functionalities.
\end{abstract}

\maketitle

\section{Introduction}
\label{section1}

Photonic devices that can function robustly for some practical applications require polarization independent responses which are immune to perturbations that can easily perturb one polarization state to another~\cite{YARIV_2006__Photonics,LIU_2005__Photonic}. For composite devices such as those consisting of periodic, quasi-periodic or disordered photonic structures~\cite{VARDENY_2013_Nat.Photonics_Optics,WIERSMA_2013_Nat.Photonics_Disordered}, usually this can be reduced to an elementary Mie scattering problem~\cite{Bohren1983_book}: its fundamental building atom needs to exhibit invariant scattering properties for different polarizations~\cite{BARRON_2009__Molecular,jahani_alldielectric_2016,KUZNETSOV_Science_optically_2016,LIU_2018_Opt.Express_Generalized}. To get rid of the polarization dependence actually constitutes a rather seminal problem in Mie theory, for which discrete spatial symmetries~\cite{BARRON_2009__Molecular,BIRSS_1964,Hopkins2013_nanoscale,FERNANDEZ-CORBATON_2013_Opt.ExpressOE_Forwarda,ZAMBRANA-PUYALTO2013Opt.Lett.,HOPKINS_2016_LaserPhotonicsRev._Circular,COLLINS_AdvancedOpticalMaterials_chirality_2017,CHEN_2020_Phys.Rev.Research_Scatteringa} and/or electromagnetic duality symmetry~\cite{FERNANDEZ-CORBATON_2013_Phys.Rev.Lett._Electromagnetica,FERNANDEZ-CORBATON_2013_Opt.ExpressOE_Forwarda,ZAMBRANA-PUYALTO2013Opt.Lett.,RAHIMZADEGAN_2018_Phys.Rev.Applied_CoreShella,YANG_2020_ACSPhotonics_Electromagnetic,YANG_2020_ArXiv200610629Phys._Scattering} can be employed to secure the scattering invariance.

A common limitation widely shared by previous studies is that the polarization independence obtained covers only some specific polarization states (generally circular or linear polarizations), occupying a rather small proportion of the whole Poincar\'{e} sphere that can represent all possible polarizations~\cite{YARIV_2006__Photonics,RAMASESHAN_1990_Curr.Sci._Poincare}. To conduct comprehensive investigations into all possible polarization states for realistic practical applications, restricting to some special polarizations is not sufficient considering the following twofold reasons: (i) Scattering invariance for some polarizations does not ensure the invariance for all polarizations throughout the whole Poincar\'{e} sphere. For example, even if the scattering properties are fully independent of linear polarizations with arbitrary orientations, the scattering variance could still emerge for states that are elliptically polarized. (ii) In realistic photonic devices, those widely explored circularly or linearly polarized states are not really absolute stable, which can be easily converted,  by inevitable structures defects or external perturbations, into more generic elliptically polarized states.

An extra problem for previous studies on symmetry protected scattering invariance is that the arguments put forward are heavily based on complicated algebraic formulas (see \textit{e.g.} Refs.~\cite{Hopkins2013_nanoscale,FERNANDEZ-CORBATON_2013_Opt.ExpressOE_Forwarda,HOPKINS_2016_LaserPhotonicsRev._Circular}), which has on one hand obscured the fundamental physical principles and on the other curbed the general interest by repelling mathematically the broader community in photonics. This echoes what is widely believed true in mathematics (and also in physics)~\cite{ATIYAH_2002_Bull.Lond.Math.Soc._MATHEMATICS,NEEDHAM_2020__Visual}: algebra is the offer from the devil to trade for our soul, stopping us from thinking geometrically and thus from grasping the underlying truth and real meaning. Basically, comprehensive revelations, justified by intuitive geometric arguments, about symmetry dictated invariant scattering for all polarizations are pressingly desired, which is exactly what we aim to present here.

In this study we show, by intuitive geometric reasoning, how to obtain symmetry-protected invariant scattering properties for arbitrary polarizations, relying solely on rotation symmetry and optical activity elimination [identical responses for left- and right-handed circularly polarized (LCP and RCP) incidnet waves]. It is discovered that for a scattering configuration of more than two-fold rotation symmetry, the two scattering channels from the LCP and RCP waves are actually decoupled: there is effectively no contribution from their cross interferences for scattering properties including extinction, absorption or total scattering. Based on this discovery, we further reveal subtle connections between scattering invariance and a hierarchy of discrete spatial symmetries: (i) Rotation symmetries ($n$-fold, $n\geq3$) result in invariance of all scattering properties for polarizations on the same latitude circle of the Poincar\'{e} sphere; (ii) Combined rotation-mirror (perpendicular to the rotation axis) or rotation-inversion symmetries lead to invariant extinctions for arbitrary polarizations, while the scattering and absorption are still variable; (iii) Combined rotation-mirror (parallel to the rotation axis) symmetry ensures invariant extinction, scattering and absorption for all polarizations covering the whole Poincar\'{e} sphere.  Since underlying those apparent spatial symmetries there are hidden functioning laws of reciprocity, parity and helicity conservation, the scattering invariance obtained is intrinsic and robust against any non-symmetry-breaking perturbations, which can potentially enrich the toolbox of optical device designs and render extra freedom for more flexible manipulations of light-matter interactions.

\begin{figure}[tp]
\centerline{\includegraphics[width=8.2cm]{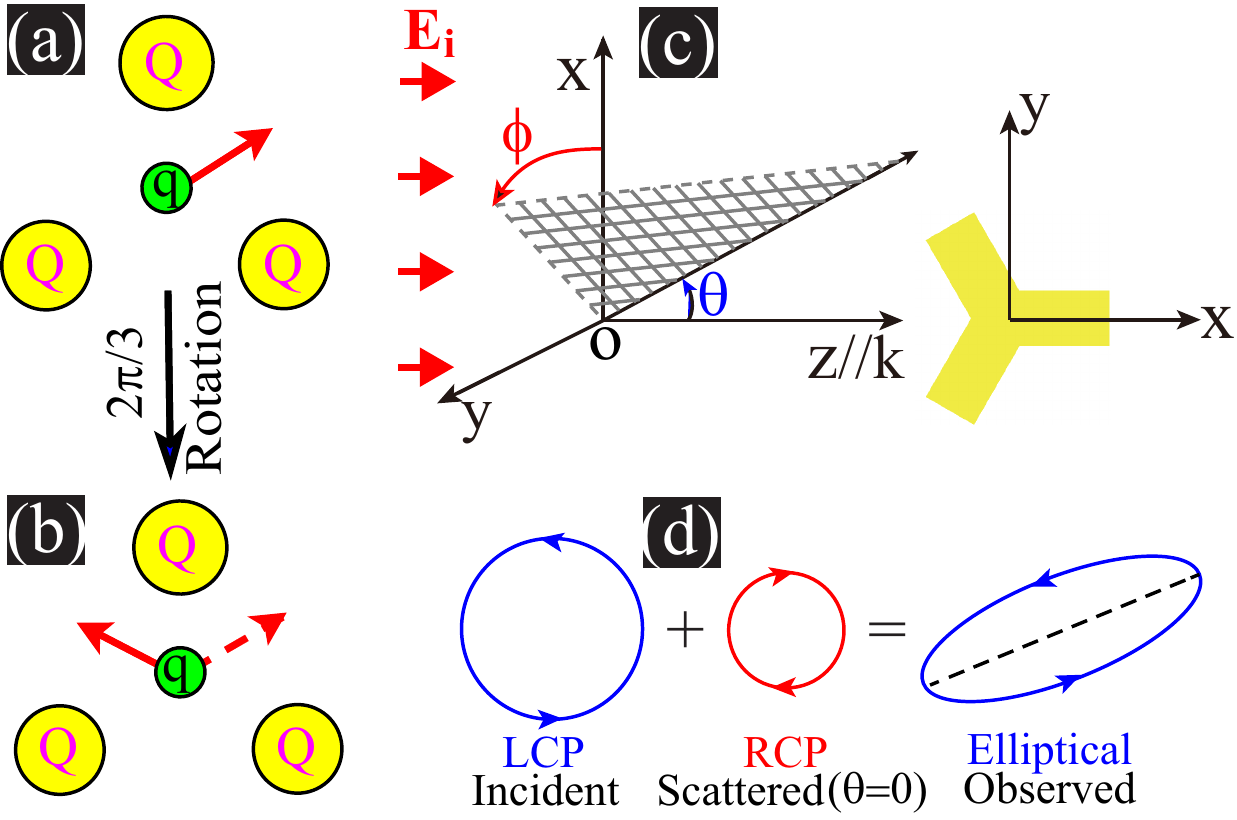}} \caption {\small (a) and (b): The geometric proof by contradiction that the electrostatic force on the central charge has to be zero with three identical charges located on the vertexes of an equilateral triangle. (c) A scattering configuration with $\rm C_3$ symmetry placed in the spherical coordinate system parameterized by polar angle $\theta$ and azimuthal angle $\phi$. (d) The geometric proof by contradiction for helicity preservation along the forward direction: \textit{e.g.} a LCP incident wave mixing with the forward scattered RCP wave would make the final observable state of elliptical polarization with preferred ellipse orientation, which is forbidden by the rotation symmetry.}\label{fig1}
\end{figure} 

\section{Helicity Preservation of forward scattering for configurations with $\mathbf{n}$-fold rotation ($\rm{\mathbf{C}_{\mathbf{n}}},~\mathbf{n\geq3}$) symmetry}
\label{section2}

In this work, our discussions of Mie scattering with incident plane waves are based on the circular basis $\mathbf{{L}}$ and $\mathbf{{R}}$,  which correspond to LCP and RCP light, respectively.  When circularly polarized (CP) waves are incident on a structure with $\rm{C_n}$ symmetry (incident direction $\textbf{k}$ is parallel to the rotation axis $l$: $\textbf{k}\|l$), the helicity is preserved along the forward direction: the forward scattered waves are not only CP but also of the same handedness as that of the incident waves.   This has already been rigorously proved by Ref.~\cite{FERNANDEZ-CORBATON_2013_Opt.ExpressOE_Forwarda}, through complicated algebraic formulas that to some extent obscure the underlying physical principles.  In this section, we provide a purely geometric formula-free proof that can directly confirm such helicity preservation, which is much simpler than that in Ref.~\cite{FERNANDEZ-CORBATON_2013_Opt.ExpressOE_Forwarda}, without sacrificing any rigor. We emphasize here that throughout this study, our arguments are restricted to $3$-fold rotation symmetry, which can be quite directly generalized to cover all scenarios of $n\geq3$.

As a first step, we turn to a seemingly unrelated problem sketched in Figs.~\ref{fig1}(a) and (b): with three identical point charges located on the vertexes of an equilateral triangle, what is the electrostatic force on an extra point charge at the triangle center? Through algebraic calculations based on the Coulomb law, we can get the answer that the force is zero. At the same time, we can reach the same conclusion through pure intuitive geometric considerations, without any detailed algebraic manipulations: (i) Assume that there is a force on the central charge as indicated by a red vector in Fig.~\ref{fig1}(a); (ii) Make a $2\pi/3$ rotation operation on the whole configuration, ending up with what is shown in Fig.~\ref{fig1}(b): both the force and the charges are rotated accordingly; (iii) Charge distributions in Figs.~\ref{fig1}(a) and (b) are identical due to the overall symmetry, requiring that the force in Fig.~\ref{fig1}(b) (dashed red vector) is the same as that in Fig.~\ref{fig1}(a); (iv) The two forces (solid and dashed red vectors) in Fig.~\ref{fig1}(b) contradict each other, unless the force is zero. This concludes our proof by contradiction.

Now we turn back to our Mie scattering problem and the scattering configuration (exhibiting $\rm{C_3}$ symmetry) within a spherical coordinate system is schematically shown in Fig.~\ref{fig1}(c).  According to Mie theory~\cite{Bohren1983_book}, what is observed in the forward direction is mixed states of both incident and forward scattered waves. For incident CP waves, let us assume that the helicity is not preserved and thus there is CP scattered components of opposite handedness along the forward direction. It is known that mixing CP waves of opposite handedness would produce elliptically polarized states with preferred orientation directions of the polarization ellipses (in terms of its semi-major or semi-minor axis): a detailed example is shown in Fig.~\ref{fig1}(d) with an incident LCP wave mixing with the forward scattered RCP wave.  According to the same arguments presented above for Figs.~\ref{fig1}(a) and (b), such a preferred orientation contradict the overall rotation symmetry of the scattering configuration with incident CP waves. When there is helicity preservation, the forward mixed state is still CP, for which there is no such contradiction as its ellipse orientation is not defined~\cite{NYE_natural_1999}). This concludes our proof by contradiction for helicity preservation along the forward direction.

We note that the same symmetry arguments can be also employed to verify the helicity flipping along the backward direction~\cite{FERNANDEZ-CORBATON_2013_Opt.ExpressOE_Forwarda}, which nevertheless is irrelevant to our following investigations and thus would not be further discussed in detail here. Moreover, there are actually subtle differences between those shown in Figs.~\ref{fig1}(a-b) and Figs.~\ref{fig1}(c-d): the electrostatic force is characterized by a vector that is variant upon the a $\pi$ rotation, and thus $\rm C_2$ symmetry is sufficient to eliminate it; while the polarization ellipse orientation is characterized by a line that is invariant upon a $\pi$ rotation, which means that more than two-fold rotation symmetry is required to guarantee the helicity preservation.

\section{Invariant scattering properties for arbitrary polarizations induced by discrete spatial symmetries}
\label{section-3}

\subsection{General theoretical analysis with intuitive geometric arguments}
\begin{figure}[tp]
\centerline{\includegraphics[width=7cm]{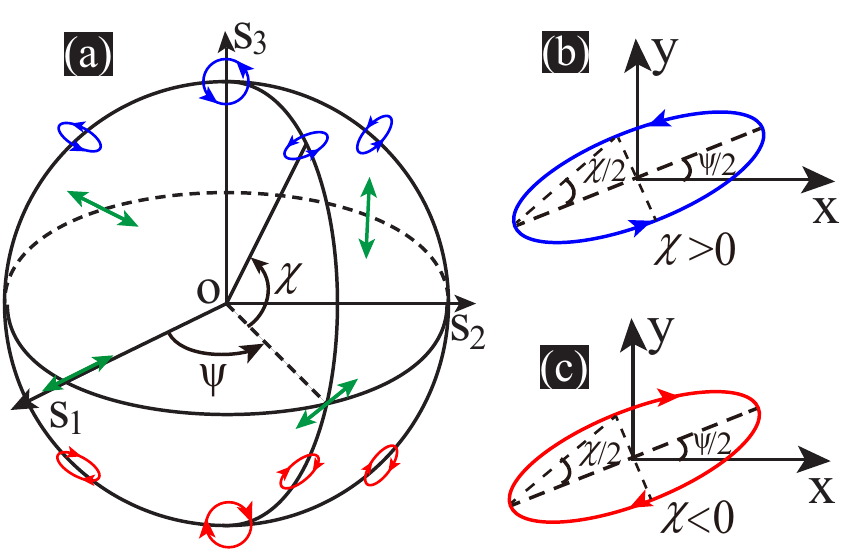}} \caption{\small (a) Representations of arbitrary polarizations by the Poincar\'{e} sphere parameterized by Stokes parameters $S_{1,2,3}$, latitude $\chi$ and longitude $\psi$. (b) and (c) The corresponding polarization ellipses are characterized by half angles of those for the Stokes vectors: $\psi/2$ indicates the ellipse orientation and $\chi/2$ describes ellipse eccentricity; polarizations of left (right) handedness locate on the northern (southern) hemisphere with $\chi>0$ ($\chi<0$); linear polarizations locate on the equator with $\chi=0$.}
\label{fig2}
\end{figure}

In this study we aim to reveal scattering invariance for arbitrary polarizations, to describe which we employ the widely adopted Poincar\'{e} sphere [characterized by Stokes vectors  ($S_{1},S_{2},S_{3}$), or location vector ($\chi$, $\psi$) in terms of latitude and longitude on a unit-sphere] as shown in Fig.~\ref{fig2}(a)~\cite{YARIV_2006__Photonics,RAMASESHAN_1990_Curr.Sci._Poincare}: latitude $\chi \in [-\pi/2, \pi/2]$ characterizes the eccentricities of the polarization ellipses,  with positive and negative $\chi$ (northern and southern hemisphere) corresponding to left and right handedness, respectively [see Figs.~\ref{fig2}(b) and (c)]; LCP and RCP waves locate respectively on the northern and southern poles, and all linear polarizations locate on the equator $S_3=\chi=0$; latitude $\psi \in [0, 2\pi]$ characterizes the orientations of the polarization ellipses in terms of the semi-major axis [see Figs.~\ref{fig2}(b) and (c)]. We emphasize here that the characterizing angles for the polarization ellipses are half of those for Stokes vectors, since the polarizations are described by 1-spinors, which are effectively the square roots of Stokes vectors~\cite{CARTAN_2012__theory,FARMELO_2011__strangest}.

An arbitrarily polarized incident wave [denoted by $\mathbf{E}_i$; located at ($\chi_i, \psi_i$) on the Poincar\'{e} sphere]  can be expressed in circular basis ($\mathbf{{L}}$, $\mathbf{{R}}$) as:
\begin{equation}
\mathbf{E}_i=\cos (\frac{\pi}{4}-\frac{\chi_i}{2})  \mathrm{e}^{-\mathrm{i} \psi_i/2} \mathbf{{L}}+\sin (\frac{\pi}{4}-\frac{\chi_i}{2}) \mathrm{e}^{\mathrm{i} \psi_i/2} \mathbf{{R}}.
\end{equation}
The scattered waves (denoted by $\mathbf{E}_s$) along different directions [characterized by ($\theta,\phi$) as shown in Fig.~\ref{fig1}(b)] are linearly related to the incident waves through the following relation~\cite{Bohren1983_book}:
\begin{equation}\begin{array}{c}
\label{scattered}
\mathbf{E}_{s}(\theta, \phi)=\hat{\mathbf{T}}(\theta, \phi) \mathbf{E}_{i}= \\
\cos (\frac{\pi}{4}-\frac{\chi_i}{2}) \mathrm{e}^{-\mathrm{i} \psi_{i} / 2} \mathbf{E}_{s}^{L}(\theta, \phi)+\sin (\frac{\pi}{4}-\frac{\chi_i}{2}) \mathrm{e}^{\mathrm{i} \psi_{i} / 2} \mathbf{E}_{s}^{R}(\theta, \phi),
\end{array}\end{equation}
where $\hat{\mathbf{T}}$ is the scattering matrix; $\mathbf{E}_{s}^{{L}}$ and $\mathbf{E}_{s}^{{R}}$ are scattered waves with incident LCP and RCP waves, respectively.

For scattering configurations with more than two-fold rotation symmetry [the scattering structure exhibits $\rm{C_n}$ ($n\geq3$) symmetry with the incident wave propagating along the rotation axis, as is the case throughout this work], the helicity preservation along the forward direction requires that $\mathbf{E}_{s}^{{L}}(\theta=0)$ and $\mathbf{E}_{s}^{{R}}(\theta=0)$ are respectively LCP and RCP waves, between which there is thus no interference. According to the optical theorem~\cite{Bohren1983_book}, the extinction is only related to interferences between incident and scattered waves along the forward direction [$\mathbf{E}_i$ and $\mathbf{E}_{s}(\theta=0)$]. Meanwhile, due to the helicity preservation, there is no cross interference between $\mathbf{{L}}$  and $\mathbf{E}_{s}^{{R}}(\theta=0)$, or between $\mathbf{{R}}$ and  $\mathbf{E}_{s}^{{L}}(\theta=0)$.  As a result, the extinction cross section for arbitrarily polarized incident waves
can be expressed as:
\begin{equation}
\label{extinction}
\rm C_{\rm{ext}}=\cos^2 (\frac{\pi}{4}-\frac{\chi_i}{2}) \rm C_{\rm{ext}}^L+\sin^2 (\frac{\pi}{4}-\frac{\chi_i}{2}) \rm C_{\rm{ext}}^R,
\end{equation}
where  $\rm {C}_{\rm{ext}}^L$ and $\rm {C}_{\rm{ext}}^R$ are extinction cross sections for incident LCP and RCP waves, respectively. According to Eq.~(\ref{extinction}), the extinction has nothing to do with $\psi_i$ (the  orientation of incident polarization ellipse) and is invariant: (i) for the incident polarizations that locate on the same circle of latitude ($\chi_i$ is constant); or (ii) for arbitrary incident polarizations when there is no extinction activity  $\rm {C}_{\rm{ext}}^L=\rm {C}_{\rm{ext}}^R$~\cite{CHEN_2020_Phys.Rev.Research_Scatteringa}.

Then we turn to scattering cross sections ($\rm C_{\rm{sca}}$) for any incident polarizations, the calculation of which appears to be more demanding than extinction since integrations of all-angular scattering intensities [$I_s(\theta, \phi)=|\mathbf{E}_{s}(\theta, \phi)|^2$] have to be implemented~\cite{Bohren1983_book}. According to  Eq.~(\ref{scattered}), the angular scattering intensity can be explicitly expressed as:
\begin{equation}
\label{intensity}
\begin{array}{c}
I_{s}(\theta, \phi)=\cos ^{2}\left(\frac{\pi}{4}-\frac{\chi_i}{2}\right) I_{s}^{L}+\sin ^{2}\left(\frac{\pi}{4}-\frac{\chi_i}{2}\right) I_{s}^{R}+ \\
\frac{1}{2}\sin(\chi_i) \cos \left(\psi_{i}\right)\left(\mathbf{E}_{s}^{* L} \cdot \mathbf{E}_{s}^{R}+\mathbf{E}_{s}^{L} \cdot \mathbf{E}_{s}^{* R}\right),
\end{array}\end{equation}
where $\ast$ means complex conjugation. Though along the forward direction $\mathbf{E}_{s}^{* L}$ and $\mathbf{E}_{s}^{R}$ are orthogonal as required by helicity preservation ($\mathbf{E}_{s}^{* L} \cdot \mathbf{E}_{s}^{R}=0$ when $\theta=0$), they are generally not orthogonal along other directions and thus the last inference term ($I_{s}^{\rm {int}}=\mathbf{E}_{s}^{* L} \cdot \mathbf{E}_{s}^{R}+\mathbf{E}_{s}^{L} \cdot \mathbf{E}_{s}^{* R}$)  cannot be directly dismissed.  This seemingly further adds to the complexities for general discussions of scattering cross sections.

Now we proceed to check in detail the scattered field distributions in terms of both phase and amplitude. For incident CP waves, the symmetry ($\rm C_3$ for the following specific discussions) of the scattering configuration ensures that $\mathbf{E}_{s}^{L, R}(\theta, \phi)$ and $\mathbf{E}_{s}^{L, R}(\theta, \phi+2m\pi/3)$ are interconnected rather than independent for $m=1,2$. It could be taken for granted that $\mathbf{E}_{s}^{L, R}(\theta, \phi)=\mathbf{E}_{s}^{L, R}(\theta, \phi+2m\pi/3)$, as they are seemingly equivalent and inter-convertible through a simple coordinate system rotation by  $2m\pi/3$ along the $z$-axis [see Fig.~\ref{fig1}(b)].  This is actually wrong, because such a coordinate rotation does not really leave the scattering configuration as it was, but would rather ultimately change it through introducing extra phase. To be specific, with the coordinate rotation, the incident CP waves would transform as follows [refer to the Feynman Lectures (Volume III, Chapter 11) for more details]~\cite{FEYNMAN_2011__Feynmanb3}:
\begin{equation}
\label{transform}
\mathbf{{L}} \rightarrow \mathrm{e}^{-2m\pi/3}\mathbf{{L}}, ~~~\mathbf{{R}} \rightarrow \mathrm{e}^{2m\pi/3}\mathbf{{R}},
\end{equation}
and the scattered fields would also transform accordingly:
\begin{equation}
\label{transform2}
\mathbf{E}_{s}^{L} \rightarrow \mathrm{e}^{-2m\pi/3}\mathbf{E}_{s}^{L}, ~~~\mathbf{E}_{s}^{R} \rightarrow \mathrm{e}^{2m\pi/3}\mathbf{E}_{s}^{R}.
\end{equation}
Equations~(\ref{intensity}) and (\ref{transform2}) lead to the following transformation for the interference term of the scattering intensity:
\begin{equation}
\label{transform3}
 I_{s}^{\rm {int}} \rightarrow \cos({4m\pi/3})I_{s}^{\rm {int}}.
\end{equation}
It is easy to conclude that the interference term would cancel each other when integrated along all scattering directions, since $1+\cos({4\pi/3})+\cos({8\pi/3})=0$. It immediately becomes clear that $\rm C_2$ symmetry does not gurantee the interference cancellation and thus there is no polarization independence, as $\cos(0)+\cos(2\pi)\neq0$. Consequently, the scattering cross section for arbitrarily polarized incident waves can be simplified as (for more than two-fold rotation symmetry):
\begin{equation}
\label{scattering}
\rm C_{\rm{sca}}=\cos^2 (\frac{\pi}{4}-\frac{\chi_i}{2}) \rm C_{\rm{sca}}^L+\sin^2 (\frac{\pi}{4}-\frac{\chi_i}{2}) \rm C_{\rm{sca}}^R,
\end{equation}
where  $\rm {C}_{\rm{sca}}^L$ and $\rm {C}_{\rm{sca}}^R$ are scattering cross sections for incident LCP and RCP waves, respectively.  Similar to Eq.~(\ref{extinction}), Eq.~(\ref{scattering}) confirms that the scattering is invariant: (i) for the incident polarizations that locate on the same circle of latitude ($\chi_i$ is constant); or (ii) for arbitrary incident polarizations when there is no scattering activity  $\rm {C}_{\rm{sca}}^L=\rm {C}_{\rm{sca}}^R$~\cite{CHEN_2020_Phys.Rev.Research_Scatteringa}. It now becomes clear that the all-angle integration actually simplifies rather than complexifies the expressions of $\rm C_{\rm{sca}}$, but for scattering configurations with rotation symmetry only.

Up to now, we have discussed only extinction and scattering, and properties of absorption can be simply deduced as the absorption cross section can be obtained through the following realtion $\rm C_{\rm{abs}}=\rm C_{\rm{ext}}-\rm C_{\rm{sca}}$, as secured by the optical theorem~\cite{Bohren1983_book}. According to Eqs.~(\ref{extinction}) and (\ref{scattering}), this leads to:
\begin{equation}
\label{absorption}
\rm C_{\rm{abs}}=\cos^2 (\frac{\pi}{4}-\frac{\chi_i}{2}) \rm C_{\rm{abs}}^L+\sin^2 (\frac{\pi}{4}-\frac{\chi_i}{2}) \rm C_{\rm{abs}}^R,
\end{equation}
where  $\rm {C}_{\rm{abs}}^L$ and $\rm {C}_{\rm{abs}}^R$ are absorption cross sections for incident LCP and RCP waves, respectively. The principles we have revealed here for $\rm C_3$  symmetry can be directly extended to any rotation symmetry $\rm C_n$ ($n>3$). We emphasize that the phase term introduced by coordinate system rotation [shown in Eq.~(\ref{transform})-(\ref{transform2})] is only observable through the interference term in Eq.~(\ref{intensity}).  When it is CP incident waves, there is no such interference and then such a phase is not observable, confirming that the arguments we presented proving the helicity preservation are still valid and not affected by the presence of such a phase.

\subsection{Invariant scattering properties induced by sole rotation symmetries}

Equations~(\ref{extinction}), (\ref{scattering}), and (\ref{absorption}) are the core results of our study, requiring the only precondition that the scattering configuration is of more than two-fold rotation symmetry. They indicate that for extinction, absorption and total scattering, there is no effective contribution from the interferences between the two scattering channels with incident LCP and RCP waves, respectively. This is to say, the two scattering channels are effectively decoupled in terms of the scattering properties we discus in this work. As a result, all scattering properties for arbitrarily polarized incident waves can be obtained by direct summations of the contributions from LCP and RCP incidences. It is clear from those equations that all scattering properties are invariant for polarizations on the same circle of latitude ($\chi_i$ is constant), a special case of which is the linear polarization ($\chi_i=0$) independence discussed in Ref.~\cite{Hopkins2013_nanoscale}.  Despite that our conclusion in this work is more general (not limited to linear polarizations), the proof we have presented here is superior: the former proof in Ref.~\cite{Hopkins2013_nanoscale} relies on the coupled dipole approximation (thus non-intrinsic) and extremely heavy algebraic manipulations (thus less accessible for general interest) that significantly obscure the physical principles; while here we provide an intrinsic symmetry-based proof assisted by intuitive geometric reasoning, which brings to the surface the key mechanisms and thus more comprehensible for the broad community in photonics.

\begin{figure}[tp]
\centerline{\includegraphics[width=8cm]{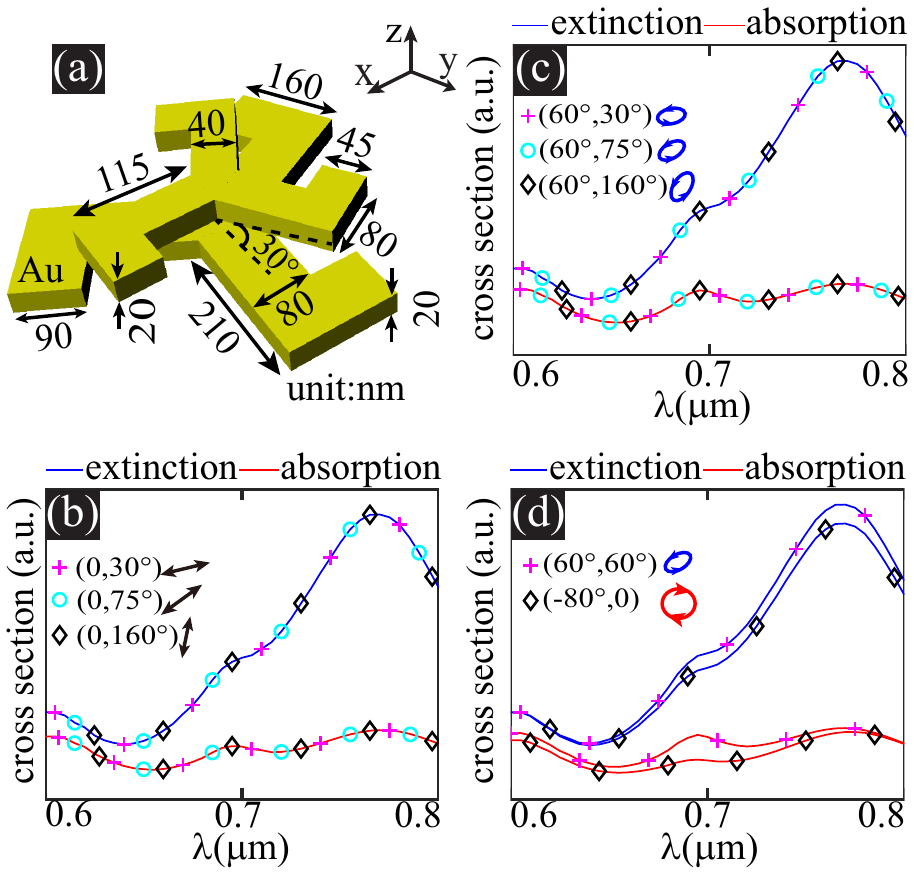}} \caption{\small (a) A  scattering configuration consisting of two touching gold particles (geometric parameters specified in the figure only) that exhibits $\rm C_3$ symmetry only. The extinction and absorption spectra are summarized in (b)-(d) for:  linear polarizations of different orientations in (b); elliptical polarizations of different orientations while on the same lattitue circle of the Poincar\'{e} sphere in (c); two randomly chosen polarizations not on the same latitude circle in (d). The labeling angles correspond to the location vector ($\chi_i$, $\psi_i$).}\label{fig3}
\end{figure}

To verify what has been claimed above, we show in Fig.~\ref{fig3}(a) a composite scattering configuration exhibiting sole $C_3$ symmetry: the two touching scatterers are made of gold, permittivity of which is taken from Ref.~\cite{Johnson1972_PRB};
the geometric parameters are specified in the figure and numerical results are obtained using COMSOL Multiphysics, as is the case throughout this work. Despite the invariance of all scattering properties (only extinction and absorption spectra are shown with respect to wavelenth $\lambda$) for linear polarizations [see Fig.~\ref{fig3}(b)], we also demonstrate in Fig.~\ref{fig3}(c) such invariance for elliptic polarizations on the same latitude circle of the Poincar\'{e} sphere ($\chi_i=60^\circ$). For polarizations not on the same latitude circle, such invariance is immediately lost, as is shown in Fig.~\ref{fig3}(d).

\subsection{Invariant scattering properties induced by combined rotation-mirror (perpendicular to the rotation axis) or rotation-inversion symmetries}

According to Eqs.~(\ref{extinction}), (\ref{scattering}), and (\ref{absorption}), to obtain invariant extinction, scattering or absorption for arbitrary polarizations (not limited to the same latitude circle on the Poincar\'{e} sphere), we only have to extinguish the corresponding scattering activities to ensure equal responses for incident LCP and RCP waves~\cite{CHEN_2020_Phys.Rev.Research_Scatteringa}: $\rm {C}_{\rm{ext,sca,abs}}^L=\rm {C}_{\rm{ext,sca,abs}}^R$, respectively.

\begin{figure}[tp]
\centerline{\includegraphics[width=8cm]{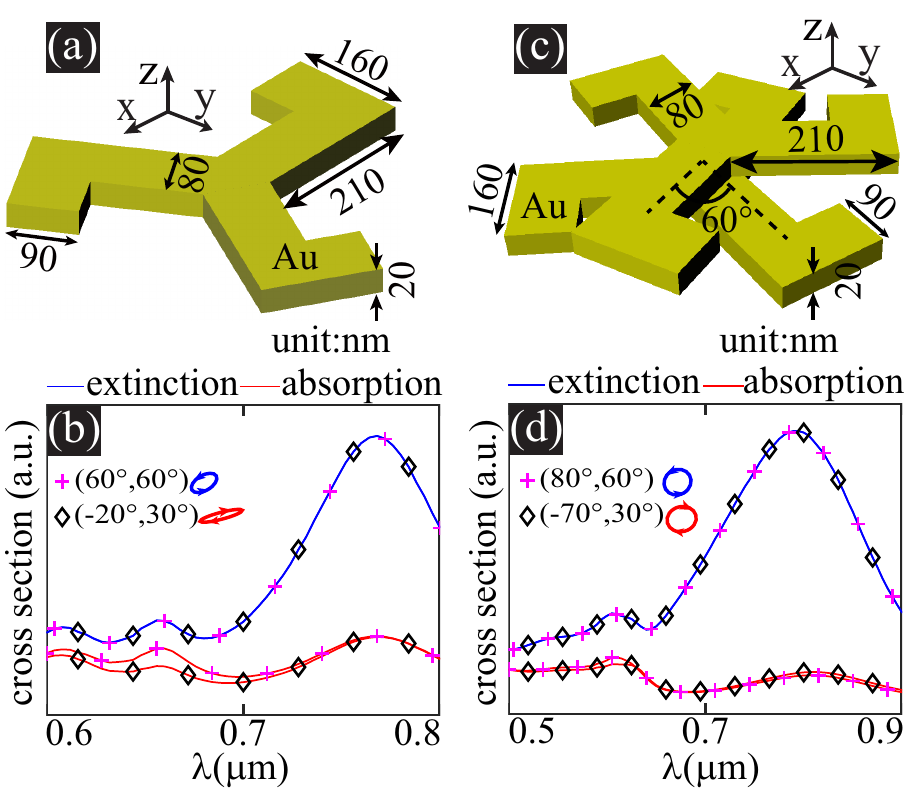}} \caption{\small Two $\rm C_3$-symmetric scattering configurations consisting of gold particles that exhibit extra perpendicular-mirror symmetry in (a) and inversion symmetry in (c) with two identical particles. The extinction and absorption spectra are summarized respectively in (b) and (d), for two randomly chosen polarizations not on the same latitude circle of the Poincar\'{e} sphere. The labeling angles correspond to the location vector ($\chi_i$, $\psi_i$).}
\label{fig4}
\end{figure}

It is recently proved that the law of reciprocity and parity conservation can intrinsically eliminate the extinction activity (but the scattering and absorption activities are still present) when there is mirror (perpendicular to the incident direction) or inversion symmetry~\cite{CHEN_2020_Phys.Rev.Research_Scatteringa}. Two such scattering configurations are shown in Figs.~\ref{fig4}(a) and (c), which besides rotation symmetry also exhibit perpendicular mirror and inversion symmetry, respectively. The scattering spectra (in terms of extinction and absorption) are shown respectively in Figs.~\ref{fig4}(b) and (d) for two randomly chosen polarizations (not on the same latitude circle).  As is clearly shown there is extinction invariance but no such invariance for absorption or scattering [the two absorption spectra in Fig.~\ref{fig4}(d) are quite close, but definitely different, which is most visible close to the wavelength $\lambda=0.6~\mu$m], since the scattering and absorption activities are not eliminated by the extra mirror or inversion symmetry~\cite{CHEN_2020_Phys.Rev.Research_Scatteringa}.

\begin{figure}[tp]
\centerline{\includegraphics[width=8cm]{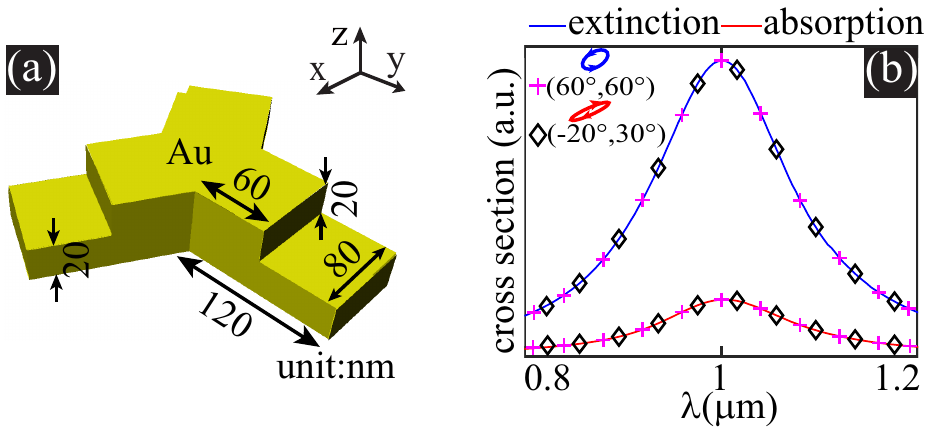}} \caption{\small  (a) A  scattering configuration consisting of two touching gold particles that exhibit both rotation and parallel-mirror ($\rm C_{3v}$) symmetry. The extinction and absorption spectra are shown in (b) for two randomly chosen polarizations not on the same latitude circle. The labeling angles correspond to the location vector ($\chi_i$, $\psi_i$).}
\label{fig5}
\end{figure}

\subsection{Invariant scattering properties induced by combined rotation-mirror (parallel to the rotation axis) symmetries}

Equations~(\ref{extinction}), (\ref{scattering}), and (\ref{absorption}) tell that to achieve invariance for all scattering properties, all the scattering activities should be eliminated simultaneously. This is possible for a configuration with extra mirror (parallel to the incident direction) symmetry, where the law of parity conservation ensures identical responses for incident LCP and RCP waves~\cite{BARRON_2009__Molecular,CHEN_2020_Phys.Rev.Research_Scatteringa}.  A scattering configuration exhibiting both rotation and parallel mirror symmetry (with broken perpendicular mirror symmetry) is shown in Fig.~\ref{fig5}(a), with the scattering and absorption spectra for two arbitrary polarizations shown in Figs.~\ref{fig5}(b), confirming the invariance of all scattering properties.

\section{Conclusions and Discussions}

In conclusion, we have revealed through intuitive geometric reasoning, how to achieve invariant scattering properties (including extinction, scattering and absorption) for arbitrary polarizations based on discrete spatial symmetries. It is discovered that sole rotational symmetries (more than 2-fold) can secure the scattering invariance for polarizations on the same latitude circle on the Poincar\'{e} sphere. To achieve invariance for all polarizations covering the whole Poincar\'{e} sphere, besides rotation symmetry we can: (i) introduce extra perpendicular mirror symmetry or inversion symmetry to produce all-polarization invariant extinctions (but not scattering or absorption); (ii) or introduce extra parallel mirror symmetry to obtain all-polarization invariant extinction, scattering and absorption.  Underlying those apparent spatial symmetries, there are functioning laws of reciprocity, helicity and parity conservation, which guarantee that the invariance obtained are intrinsic and robust against any symmetry-preserving structure defects or perturbations.

Here in this study, we have confined our discussions to fully-polarized incident waves on the Poincar\'{e} sphere, but neglect those unpolarized and partially polarized states that are within the Poincar\'{e} sphere.  Since our core results shown in Eqs~(\ref{extinction}), (\ref{scattering}), and (\ref{absorption}) have nothing to do with the cross interference term or relative phase difference between two different scattering channels, and thus all equations are valid for states within the Poincar\'{e} sphere. It is worth noting that for unpolarized incident light, the interference term is automatically cancelled and thus the validity of Eqs~(\ref{extinction}), (\ref{scattering}), and (\ref{absorption}) does not reside on the rotation symmetry of the scattering configuration anymore. It is revealed that to obtain invariance for arbitrary polarizations, the absence of optical activities is crucial, which can be either intrinsic (protected by symmetry and thus is broadband as shown in this work) or accidental (activities are only eliminated at some specific wavelengths). The principles we have revealed in this work and also the novel approaches we have employed in our intuitive geometric reasoning can shed new light on not only optical device designs, but also on fundamental explorations in photonics where the light-matter interactions are dictated by symmetry.

\section*{acknowledgement}
We acknowledge the financial support from National Natural Science
Foundation of China (Grant No. 11874026, 11404403 and 11874426), and the Outstanding Young Researcher Scheme of National University of Defense Technology.

Q. Yang and W. Chen contributed equally to this work.


\end{document}